\documentclass[a4paper,11pt]{article}
\pdfoutput=1

\usepackage{jcappub}
\usepackage[utf8]{inputenc} 
\usepackage{amsfonts,amssymb,mathrsfs,amsmath,esint,bm}
\allowdisplaybreaks

\graphicspath{{./Figures_Draft/}
{../../../Research/Shared_Open-Projects/SGWB-Discrimination/Github/sgwb/sgwb/}}
\usepackage{latexsym}
\usepackage{graphicx}
\usepackage[dvipsnames]{xcolor}
\usepackage{booktabs}

\newcommand{\gw}{\textsc{gw}}
\newcommand{\OGW}{\Omega_\gw}
\newcommand{\bbh}{\textsc{(bbh)}}
\newcommand{\bbhe}{{\textsc{(bbh,$e$\hspace{-.5pt}$>$\hspace{-.5pt}0)}}}
\newcommand{\bwd}{\textsc{(bwd)}}
\newcommand{\pt}{\textsc{(pt)}}
\newcommand{\GeV}{\,\text{GeV}}
\newcommand{\mHz}{\,\text{mHz}}
\newcommand{\Hz}{\,\text{Hz}}
\newcommand{\kHz}{\,\text{kHz}}
\newcommand{\Msun}{M_\odot}
\renewcommand{\d}{\mathrm{d}}
\newcommand{\fFS}{f_\textsc{fs}}
\newcommand{\Hs}{H_*}
\newcommand{\Ts}{T_\textsc{pt}}
\newcommand{\Sh}{S_h}
\newcommand{\Sp}{S_\text{prim}}
\newcommand{\Sa}{S_\text{astro}}
\newcommand{\Aa}{A_\text{astro}}
\newcommand{\avAa}{\overline{A}_\text{astro}}
\newcommand{\sa}{\sigma_\text{astro}}
\newcommand{\s}{\bm{s}}
\renewcommand{\a}{\bm{a}}
\newcommand{\N}{\mathcal{N}}
\newcommand{\R}{\mathcal{R}}
\newcommand{\SNR}{\textsc{snr}}

\newcommand{\DR}[1]{}
\newcommand{\cita}[1]{}

\title{Precision cosmology with primordial GW backgrounds in presence of astrophysical foregrounds}
\author[a,1]{D.~Racco}
\author[b]{D.~Poletti} 

\affiliation[a]{Stanford Institute for Theoretical Physics, Stanford University, Stanford, CA 94305, USA}
\affiliation[b]{University of Milano-Bicocca, Piazza della Scienza, 3 - 20126 Milano (MI), Italy.}

\emailAdd{dracco@stanford.edu}

\abstract{
The era of Gravitational-Wave (GW) astronomy will grant the detection of the astrophysical GW background from unresolved mergers of binary black holes, and the prospect of probing the presence of primordial GW backgrounds. 
In particular, the low-frequency tail of the GW spectrum for causally-generated primordial signals (like a phase transition) offers an excellent opportunity to measure unambiguously cosmological parameters as the equation of state of the universe, or free-streaming particles at epochs well before recombination.
We discuss whether this programme is jeopardised by the uncertainties on the astrophysical GW foregrounds that coexist with a primordial background.
We detail the motivated assumptions under which the astrophysical foregrounds can be assumed to be known in shape, and only uncertain in their normalisation. 
In this case, the sensitivity to a primordial signal can be computed  by a simple and numerically agile procedure, where the optimal filter function subtracts the components of the astrophysical foreground that are close in spectral shape to the signal.
We show that the degradation of the sensitivity to the signal in presence of astrophysical foregrounds is limited to a factor of a few, and only around the frequencies where the signal is closer to the foregrounds.
Our results highlight the importance of modelling the contributions of eccentric or intermediate-mass black hole binaries to the GW background, to consolidate the prospects to perform precision cosmology with primordial GW backgrounds.
}

\begin{document}

\maketitle

\section{Introduction}

The first direct observation \cite{LIGOScientific:2016aoc} of Gravitational Waves (GWs) performed in 2015 by the interferometers of LIGO, whose measurements were joined shortly after by the companion VIRGO and later by KAGRA, marked the onset of the GW astronomy era. 
After that first detection of the merger of a black hole binary (BBH), the catalogues released by the LVK consortium after the run O3 of observation \cite{LIGOScientific:2020ibl, LIGOScientific:2021djp} collect almost 90 compact binary coalescences, predominantly BBH mergers together with two binary neutron star (BNS) mergers (including the first multi-messenger detection of a merger \cite{LIGOScientific:2017vwq, LIGOScientific:2017ync, LIGOScientific:2017zic}), and a few candidate BH-NS merger.
Direct observations of GWs offer a completely new window to explore the Universe, which has spurred many directions to interpret the incoming wealth of data for a variety of scientific programs in astrophysics, cosmology and fundamental physics \cite{Maggiore:2006uy, Barack:2018yly, Sathyaprakash:2019yqt}.

Besides the observation of individually resolved mergers, another frontier lies ahead in GW astronomy: the detection of the stochastic GW background (SGWB) filling our Universe (see \cite{Caprini:2018mtu,Renzini:2022alw} for recent reviews).
This background is generated by the stochastic superposition of multiple GWs travelling cosmological distances and originates from various contributions. 
A class of GW backgrounds is of astrophysical nature, and consists of the incoherent superposition of the GWs generated by all the binaries that have merged throughout the cosmic history, including BBH, BNS (which are subleading in terms of SGWB), and White Dwarfs binaries (BWD).
\DR{other bkgs, like core-collapse SNe etc?}

Along with this background, that will be measured \cite{Allen:1997ad} by the end of the mission of LVK \cite{KAGRA:2021kbb} or at the latest by future ground-based (CE, ET) \cite{Punturo:2010zz, Reitze:2019iox, Sharma:2020btq, Kalogera:2021bya} and space-based (LISA, BBO, MAGIS, DECIGO) \cite{LISACosmologyWorkingGroup:2022jok, Harry:2006fi, Graham:2017pmn, Kawamura:2011zz} experiments, there could lie a primordial signal generated in the early cosmological history \cite{Starobinsky:1979ty, Grishchuk:1993te,Turner:1996ck, Maggiore:1999vm, Buonanno:2003th, Bartolo:2016ami, Guzzetti:2016mkm}.
Such a discovery would offer to us the farthest signal that we have ever detected, way beyond what is presently visible in the Cosmic Microwave Background (CMB). 
GWs can travel to us (almost \cite{Bartolo:2019yeu}) unperturbed, offering us with a pristine glance into the earliest cosmological epochs.
For these reasons, a primordial GW background can be a powerful tool to probe the cosmological evolution and the particle content of the Universe at primordial times, since shortly after the generation of the GWs until today.
The power spectrum of the SGWB offers the tantalising prospect of performing precision measurements for cosmology~\cite{%
Seto:2003kc,
Boyle:2005se,
Watanabe:2006qe,
Boyle:2007zx,
Jinno:2012xb,
Caprini:2015zlo,
Geller:2018mwu,
Saikawa:2018rcs,
Cui:2018rwi,
Caldwell:2018giq,
DEramo:2019tit,
Figueroa:2019paj,
Auclair:2019wcv,
Chang:2019mza,
Caprini:2019egz,
Gouttenoire:2019kij,
Gouttenoire:2019rtn,
Blasi:2020wpy,
Domenech:2020kqm,
Allahverdi:2020bys}.
To put this program in practice, we need to know what was the original spectral shape of the SGWB at formation, to detect the subsequent modifications throughout the cosmological history. 
This is not the typical case, on general grounds. 
A remarkable exception comes from the class of GW backgrounds that are generated from a phenomenon happening within a finite time and locally in each spatial patch, the prototypical example being a phase transition (PT).
In that case, the low-frequency tail of the spectrum, corresponding to wavelengths larger than the Hubble radius (or horizon) at the time of GW generation, is constrained by causality arguments to have a power-law dependence, whose tilt informs us of the expansion history of the Universe and the particle content of the Universe \cite{Hook:2020phx, Brzeminski:2022haa,Loverde:2022wih}.
The physical origin of this result can be explained in simple terms \cite{Hook:2020phx} by considering the evolution of modes on scales larger than the correlation length of the source of GWs. 
The result is that the spectrum at low frequencies (or causality-limited tail) is independent of the generation mechanism, and can only be affected by (\textit{i}) the expansion history, which affects the redshift of waves inside the horizon and the evolution of super-horizon modes, and (\textit{ii}) the propagation of GWs, which do not travel as free waves if there is anisotropic stress in the cosmic fluid. 
The latter can notably be sourced by free-streaming particles, providing us with the unique opportunity of detecting their presence before the epoch of recombination. \DR{check other sources of $\pi_{ij}$}

The question to which we wish to provide an answer in this paper is the following one:
In the simultaneous presence of astrophysical GW backgrounds and of a primordial GW background of causal origin (e.g.\,from a PT), how well can we perform precision cosmology through the low-frequency part of the primordial signal?
In order to discriminate these backgrounds (we refer to the astrophysical ones as \textit{foregrounds}) we consider the analysis of the global average signal in frequency space, analogously to the measurement by COBE of the black-body spectrum of the CMB \cite{Mather:1990tfx}. 
This path, that has been mostly explored so far in the literature \cite{%
Sachdev:2020bkk,
Flauger:2020qyi,
Boileau:2020rpg,
Boileau:2021sni,
Gowling:2021gcy,
Boileau:2021gbr,
Lewicki:2021kmu,
Boileau:2022ter%
}, is not the only option: more refined measurements of the SGWB will detect angular anisotropies to some degree of accuracy, enabling us to use this information to further discriminate primordial and astrophysical SGWBs \cite{Bartolo:2019yeu, Capurri:2021zli, LISACosmologyWorkingGroup:2022kbp}.

In this paper, we focus on the range of frequencies ($10^{-4}-10^{-1}$ Hz) that will be measured by the space-based experiment LISA. Among the reasons for this, this experiment is at an advanced stage among various proposals (after the successful performance of the LISA pathfinder), and its GW-frequency range will provide an interesting window on primordial signals from PTs at energies not far from the present collider reach. 
Finally, this range is particularly illustrative for our method thanks to the simultaneous presence of various astrophysical foregrounds: GWs from BWD mergers in our Galaxy will be detectable and, although a significant fraction will be resolvable, there will be an irreducible unresolved component in the LISA frequency band.
Ground-based experiments, like ET or CE, will measure the peak of the BBH foreground, which has more significant modelling uncertainties; in the frequency range of LISA, instead, it is a good approximation to consider its spectral shape to be known (as we detail in the following together with all the caveats and assumptions).

We perform a Fisher matrix forecast of the sensitivity of LISA to features in the causality-limited tail (i.e.~low-frequency range) of a primordial SGWB, while marginalising over astrophysical foregrounds.
Previous studies have performed related analyses, looking at the sensitivity of LISA to a primordial SGWB that follows a power-law spectrum (in particular, flat for inflation or cosmic strings) \cite{Sachdev:2020bkk, Boileau:2020rpg, Boileau:2021sni, Boileau:2021gbr, Lewicki:2021kmu} or from a PT (with a focus on the model-dependent high-frequency part of the spectrum) \cite{Gowling:2021gcy, Boileau:2022ter}, in presence of astrophysical foregrounds from WD mergers and a power-law SGWB from unresolved BBHs.
An interesting study \cite{Pan:2019uyn} shows the potential for BBH foreground cleaning at LISA by exploiting the measurements at higher frequencies in ground-based experiments.
We differ here from previous studies in two aspects. First, we assess more carefully the spectral shape expected for astrophysical foregrounds, discussing in particular the validity of the extrapolation to low frequencies of the slope of the SGWB from BBH (Sec.~\ref{sec:BBH}, \ref{sec:BBH ecc}). Secondly, we make advantage of the statistical method introduced in \cite{Poletti:2021ytu} to optimally filter the primordial signal in presence of foregrounds with known shape and unknown normalisation.
This method is particularly suitable for our case, because the spectral shape of the SGWB from BH binaries can be treated as fixed up to a good accuracy in the frequency range of interest, and deviations (due in particular to eccentric binaries) can be shown to have a moderate impact under the assumptions that we discuss.
Also the SGWB from WD binaries, after the subtraction of resolvable mergers, will be measured by LISA and we will be able to treat it as a known source in the frequency range where LISA is most sensitive.

We begin in Sec.~\ref{sec:astro fg} with the discussion of the various astrophysical foregrounds which are relevant to the range of LISA, while in Sec.~\ref{sec:prim} we review the class of primordial signals (i.e.~whose production is causality-limited, like PT) that allows us to perform precision cosmology.
We discuss in Sec.~\ref{sec:method} the statistical treatment that we use to perform the analysis.
In Sec.~\ref{sec:results} we show our results for the sensitivity to cosmological observables in a primordial SGWB in presence of astrophysical foregrounds, and in Sec.~\ref{sec:conclusions} we summarise our conclusions.

\section{Astrophysical backgrounds of gravitational waves}
\label{sec:astro fg}
In this section, we discuss the most relevant astrophysical foregrounds for the detection of a primordial SGWB, that are the unresolved mergers of binary compact objects, considering in particular the frequency range of LISA.
We begin by reviewing the contribution from BBHs in Sec.~\ref{sec:BBH} to understand the spectrum of the associated SGWB, and we elaborate on the contribution from eccentric binaries (which is a novelty of this paper) in Sec.~\ref{sec:BBH ecc}.
Then we discuss BWDs, which are another major contribution for the LISA frequency range, in Sec.~\ref{sec:BWD}.
The SGWBs expected from BNS and NS-BH binaries are subleading to the SGWB from BBHs and display a completely analogous spectrum (see e.g.~\cite{LIGOScientific:2021psn,Perigois:2021ovr}), so we do not discuss them further.

\subsection{Black hole binaries (BBH)}
\label{sec:BBH}
As mentioned in the introduction, the detection of the SGWB from unresolved BBHs is one of the scientific goals of the LVK collaboration at their design sensitivity.
At present, the collaboration has not detected a signal, neither in  searches for an isotropic \cite{KAGRA:2021kbb} nor anisotropic \cite{KAGRA:2021mth} SGWB.
The present upper limit after O3 for the isotropic SGWB, parameterised as a power law, is around $\OGW^\bbh(25\Hz)\lesssim 10^{-8}$, with small variations as a function of the power-law tilt and the prior on $\OGW$.
The expected signal lies one order of magnitude below the present sensitivity: a recent assessment by the LVK collaboration, based on the merger rates measured through O3, infers \cite{LIGOScientific:2021psn}
\begin{equation}
\label{eq:bbh norm}
\OGW^\bbh(25\Hz) = 5.0^{+1.4}_{-1.8} \cdot 10^{-10}\,,
\end{equation}
and we use this central value for our analysis.

The calculation of the SGWB from unresolved BBHs is affected by many astrophysical uncertainties, including the star formation rate and the average metallicity as a function of redshift, the BH formation rate, and subsequently the BH binary formation rate, which in turn depends on a variety of possible formation channels \cite{Mapelli:2021taw}\cita{altro?}.
Remarkably, many of these astrophysical effects mostly manifest themselves as an uncertainty on the \textit{normalisation} of the signal.
This is particularly important for us with respect to the knowledge of the spectral shape of the SGWB. 
The measurement of $\OGW^\bbh(f)$ in the final stages of LVK will provide us with observational input for the calibration of the astrophysical uncertainties behind the BBH merger rate. 
Ground-based experiments are limited by seismic noise to measure this SGWB at frequencies $f\sim 10-10^2\Hz$, just below its peak around 100-500 Hz \cite{LIGOScientific:2021psn}.
In order to make use of this measurement to deduce the SGWB expected at lower frequencies, we need to extrapolate from the measured amplitude of the BBH foreground with the knowledge of the spectral shape \cite{Perigois:2020ymr}\cita{altri su previsione AGWB}.

The prediction from General Relativity (GR) of the spectral shape of the superposition of GWs from BBH mergers is a power law with spectral tilt of $2/3$, with some specifications that we discuss below.
This scaling can be justified as follows. 
By highlighting the main frequency dependencies, and remembering that the energy density of GWs is proportional to $f^2$,
\begin{equation}
\label{eq:OGW}
\OGW(f)\equiv \frac{1}{\rho_\text{cr}} \frac{\d \rho_\gw(f)}{\d \ln f} \propto f \frac{\d \rho_\gw(f)}{\d f} \propto f \cdot f^2 \mathcal A^2(f),
\end{equation}
where $\rho_\text{cr}=3H_0^2 M_\textsc{p}^2$ is the critical energy density, $\mathcal A(f)$ is the amplitude of the emitted GW in Fourier space, and for a circular orbit with non-spinning, point-like sources is (see e.g.~\cite[Problem 4.1]{Maggiore:2007ulw} for a derivation)
\begin{equation}
\label{eq:OGW 2/3}
\mathcal A(f) \propto f^{-\tfrac76} \quad \Rightarrow \quad \frac{\d \OGW(f)}{\d \ln f} \propto f^{2/3} \,.
\end{equation}
A few qualifications to this equation are in order.
\begin{enumerate}
\item This result is valid before the source frequency approaches 
\begin{equation}
\label{eq:f_ISCO}
f_{s,\textsc{isco}} \simeq 2.2 \kHz \frac{\Msun}{m_1+m_2}
\end{equation}
($m_{1,2}$ being the mergers' masses) \cite{Maggiore:2007ulw}, at which the merging objects plunge into each other and strong-field effects cannot be neglected. 
The redshift integral that is understood in Eq.~\eqref{eq:OGW} includes the BBH merger rate as a function of BH masses, which determine $f_{s,\textsc{isco}}$ and hence the frequency range up to which Eq.~\eqref{eq:OGW 2/3} is accurate. 
With the modelling and data presently available to the LVK collaboration \cite{LIGOScientific:2021psn}, the BH mass function is dominated by the peak at low masses ($\sim\mathcal O(1-10) \Msun$), so that $\OGW(f)$ is mostly sourced by the mergers of these light BHs. 
If instead some subpopulation at larger masses turned out to give a BBH merger rate that is large enough to overcome the $\OGW$ emitted by lighter BBHs, the overall spectrum of $\OGW(f)$ would deviate from $f^{2/3}$ around the merging frequency of that heavier subpopulation.
We assume that this is not the case (although we comment in Sec.~\ref{sec:BBH ecc} on the class of Extreme-Mass-Ratio Inspirals, or EMRI), and that most of the SGWB from BBHs (at least down to $f\sim\mHz$) is due to GWs emitted in the inspiral phase of the mergers of light BBHs.

\item Eq.~\eqref{eq:OGW 2/3} is obtained in flat background, for non-spinning BHs, and neglecting the secondary mass. 
It is possible to take into account all of these effects systematically by means of a perturbative expansion of GR corrections, known as Post-Newtonian (PN) expansion (see \cite{Blanchet:2013haa} and \cite[Chapter 5]{Maggiore:2007ulw} for reviews on the topic).
The expansion parameter is 
\begin{equation}
\label{eq:PN parameter}
x \equiv \left( \frac{G(m_1+m_2) 2\pi f_s}{c^3}\right) = \frac{R_S}{2r} \approx \frac{v^2}{c^2} \,,
\end{equation}
where $f_s$ is the source rotation frequency, $R_S=2G(m_1+m_2)/c^2$ is the Schwarzschild radius of the source, and $v$ is the velocity of the inspiralling bodies.
In order to detect the merger signal out of the large noise, we need to track the number of GW cycles at least up to $\mathcal O(1)$. 
This requires a very accurate determination of the inspiralling phase, including corrections up to the order $x^{7/2}$ or 3.5PN (see \cite{Blanchet:2013haa} for state-of-the-art results in the field). 
For the sake of discussing the leading corrections to the GW amplitude in Eq.~\eqref{eq:OGW 2/3}, it is enough the consider the first leading correction.
We can understand from Eq.~\eqref{eq:PN parameter} that the leading order in the PN expansion amounts to neglect GR corrections up to when the inspiralling bodies approach each other to distances close to $R_S$, and reach relativistic speeds. 
If, for the frequency range that we are interested in (in our case, down to $f\sim\mHz$), the leading contributions to $\OGW^\bbh(f)$ come from light BBHs far from merging (see the discussion at point 1), that is $f\ll f_{s,\textsc{isco}}$, we see that $x\sim 0$ to a good accuracy, and the PN corrections to $\mathcal A(f)$ are negligible.
In this regime (also called ``restricted'' PN \cite{Maggiore:2007ulw}) where the PN corrections to the $\mathcal A(f)$ are neglected, the corrections to Eq.~\eqref{eq:OGW 2/3} due to $m_2>0$, $m_1\neq m_2$ and spins $\vec S_1, \vec S_2 \neq 0$ (which can be found in \cite{Blanchet:2013haa}) are irrelevant.
There is only one orbital parameter which is still relevant to $\OGW^\bbh$, the eccentricity $e$ (see point 3).

\item Eq.~\eqref{eq:OGW 2/3} is valid for circular orbits with $e=0$.
This is a good assumption for all the BBHs that are born with small eccentricity, and also for any BBH after some time through the inspiral phase, because the GW emission tends to quickly reduce the eccentricity \cite{Peters:1963ux,Peters:1964zz}.
If a sizable fraction of BBHs arises from a formation channel which leads to large $e>0$ at formation, then its GW emissions (for about a decade in frequency through the inspiral) are affected by the eccentricity.
We dedicate Sec.~\ref{sec:BBH ecc} to discuss how we take this effect into account.
\end{enumerate}

In summary, we have justified why it is a good approximation to treat the SGWB from unresolved BBHs as a power law $\OGW^\bbh(f) \propto f^{2/3}$ below its peak around $10-100 \Hz$.
The class of eccentric binaries, whose amplitude evolution differs from non-eccentric binaries, deserves a more careful assessment, that we discuss in Sec.~\ref{sec:BBH ecc}, but seems not to affect significantly this prediction down to the frequency range of LISA. 

A separate effect that could be relevant for this discussion arises when we consider the subtraction from $\OGW^\bbh(f)$ of single binaries whose waveform can be individually identified, as studied e.g.\ by \cite{Sachdev:2020bkk,Zhou:2022nmt}. 
The reconstruction of the BBH parameters for these resolvable contributions is inevitably imperfect, so that there is a residual difference between the actual waveform and the reconstructed one. 
The superposition of these residuals should still follow  the $f^{2/3}$ power law, because if both the actual GW amplitude and the reconstructed one follow Eq.~\eqref{eq:OGW 2/3}, then their difference should too%
\footnote{We notice that in \cite{Sachdev:2020bkk} the residual $\OGW^\bbh$ after subtraction deviates from $f^{2/3}$, but this effect disappears when accounting for the reconstruction error on more parameters than just chirp mass $\mathcal M_z$, coalescence phase $\phi_c$ and coalescence time $t_c$, as apparent from Figs.~6 and 7 of \cite{Zhou:2022nmt}.}. 
For this reason, the subtraction procedure should not affect our discussion.

\subsection{Eccentric black hole binaries}
\label{sec:BBH ecc}
As illustrated in the previous section, the most significant contribution that could imply deviations of $\OGW^\bbh(f)\propto f^{2/3}$ at $f\sim\mHz$ are the GWs emitted from BBHs with large $e$ at formation.
There are known astrophysical environments leading to the formation of BBHs with large eccentricity (see \cite{Mapelli:2021taw} for a review of BBH formation channels).
 
Quieter environments in the gravitational field of the galaxy allow binary stellar systems to evolve via the formation of a common envelope of stellar material towards the end of their life, whose friction brings them close enough that they form individual black holes that merge within a Hubble time. 
These \textit{isolated} BBHs typically display $e\approx 0$ at formation.

Different formation channels lead to the generation of a BBH as a result of dynamical interactions with other stars or BHs, and the eccentricity of the merging binary can be of order 1. 
Such \textit{dynamical} BBHs can arise in dense stellar systems, such as star clusters, where 3-body interactions are frequent.%
\footnote{A known phenomenon leading to eccentric BBHs out of 3-body systems are Lidov-Kozai oscillations \cite{Lidov:1962,Kozai:1962zz,Antonini:2012ad}.
In these hierarchical systems, a light secondary mass orbits around the heavy primary (inner orbit), and a far heavy perturber orbits around the centre-of-mass of primary and secondary (outer orbit). 
Inner and outer orbits are inclined by an inclination angle $i$, and the eccentricity of the inner orbit is $e$.
The conservation of the angular momentum $\vec L_\text{secondary}$ in the direction of $\vec L_\text{perturber}$ implies that $\sqrt{1-e^2}\cos i$ is constant.
On very long (Kozai) timescales, $e$ and $i$ oscillate: when the inner orbit becomes more inclined (compared to the outer orbit plane), its eccentricity increases. 
If dissipative processes (like GW emission
) reduce the inner orbit radius faster than the Kozai timescale, then the inner system decouples from the perturber and undergoes merging.
}
One of the scientific goals of the GW community is to infer the fractions contributing to the population of merging BBHs for each formation channel, on the basis of the LVK data. 
These fractions cannot be discriminated from $e$ at merging, given that the orbits have already circularised, but from the study of of mass, spin and redshift distributions and the comparison with simulated catalogues.
Future data and developments of semi-analytical modelling of BBH-forming environments will allow to refine these analyses, and a growing body of studies in the literature (with very few exceptions \cite{Galaudage:2021rkt}) finds that a fraction $\mathcal O(0.1-1)$ of the BBHs measured by LVK is of dynamical origin \cite{Ford:2021kcw, Bavera:2021wmw, Sedda:2021vjh, Franciolini:2022iaa, Stevenson:2022djs}.

The effect of eccentric binaries on the SGWB from unresolved binaries is the following (see \cite{Peters:1963ux,Peters:1964zz} for the original analysis).
The Fourier spectrum of the GWs emitted by an eccentric binary is continuous, rather than discrete as a circular binary. The peak frequency $f_p$, corresponding to the separation of the mergers at periastron, %
scales with time as $\dot f_p \sim f_p^{11/3} \sqrt{1-e^2} \overset{e\to 1}\to 0$. 
A BBH spends more time emitting GWs at $f_p$ if the orbit is eccentric, and more BBHs accumulate in that frequency bin, potentially distorting the power-law shape of the SGWB (see \cite{Fang:2019dnh} for a recent analysis).

In order to assess the net impact of eccentric binaries on the spectral shape of $\OGW^\bbh$, the way forward is a comprehensive modelling of many astrophysical variables determining the BBH formation rate.
These include in particular metallicity, star formation rate, stellar binary formation rate, efficiency in the ejection of stellar material in the common envelope, and 3-body dynamics in the environment.
Many groups are facing this programme with semi-numerical approaches (see e.g.~\cite{Perigois:2020ymr,Mapelli:2021gyv,Perigois:2021ovr} and references therein)\cita{altri?}, and we can make use of two recent studies to get a concrete expectation of the size of the residual SGWB of eccentric BBH with respect to the power-law $f^{2/3}$:
\begin{equation}
\OGW^\bbhe(f) \equiv \OGW^\bbh(f) - \left(\OGW^\bbh(f)\Big|_{e=0}\right) \,,
\end{equation}
where $\OGW^\bbh(f)|_{e=0}\propto f^{2/3}$ is defined as the SGWB computed by fictitiously ignoring the eccentricity effects on the inspiral.
A first ingredient to assess this is an estimate of the fraction of observed BBH mergers coming from dynamic environments that are known to produce eccentric BBHs. 
The recent state-of-the-art analysis of \cite{Mapelli:2021gyv} quantifies, for a few benchmarks choices for the astrophysical inputs, the fractions of mergers from isolated binaries, and from binaries in nuclear star clusters (NSC), globular star clusters (GSC) and young star clusters (YSC).
Starting from their population synthesis pipeline, they infer the expected distributions for the BBH merger parameters, which are then compared to data in order to select the more realistic astrophysical benchmarks.
We make use of the average fraction of mergers from YSCs from \cite{Mapelli:2021gyv} to properly weigh the residual SGWB $\OGW^\bbhe$ obtained in the follow-up study \cite{Perigois:2021ovr}, which simulates a catalogue of compact objects%
\footnote{In the findings of \cite{Perigois:2021ovr}, only BBHs end up being relevant in terms of SGWB from unresolved mergers, while BNS and NS-BH binaries give a sub-leading contribution.}
in YSCs and computes the SGWB from the binaries merging within a Hubble time.
The result of our estimate is (for a wide frequency range $f\gtrsim \mu\textrm{Hz}$)
\begin{align}
\OGW^\bbhe(f) \lesssim 10^{-14} & \qquad \text{(for BBHs in YSC \cite{Perigois:2021ovr})} \,, \\
\OGW^\bbhe(f) \lesssim 10^{-13} & \qquad \text{(for BBHs in NSC, GSC, YSC \cite{Mapelli:2021gyv}\&\cite{Perigois:2021ovr})} \,.
\end{align}
In the remainder of our analysis, we conservatively assume that future refinements of population synthesis studies (informed by new data from resolved mergers at LVK), and the inclusion of additional galactic environments in these analyses, will at most increase the estimate of this SGWB by a factor of 10:
\begin{equation}
\label{eq:BBH ecc}
\OGW^\bbhe(f) \lesssim 10^{-12} \qquad \text{(conservative; this work)} \,.
\end{equation}
We understand that this procedure, admittedly approximate, has the goal of providing us with an educated guess of the final impact of eccentric binaries on the $f^{2/3}$ power-law shape, in order to justify our treatment of $\OGW^\bbh$ in the following (see Sec.~\ref{sec:method}).
As we elaborate further later, we actually believe that progress in the direction of estimating the impact of eccentric binaries on the SGWB expected from BBHs will be very valuable to improve our sensitivity to primordial SGWBs.

Before concluding this section, we would like to comment on another possible source of uncertainty.
A class of BBHs which could affect the assumption (1) that we discussed after Eq.~\eqref{eq:OGW 2/3} are Extreme Mass-Ratio Inspirals (EMRI), composed by an intermediate-mass and a stellar-mass BHs. 
The EMRIs that are visible with LISA have a heavy progenitor around $10^{5-7} M_\odot$, which falls in the mass range of the BHs typically hosted at galactic centres. 
If the merger rates of EMRIs turned out to be comparable to the ones of stellar-mass BHs, their $f_{s,\textsc{isco}}$ would lie around the peak sensitivity of LISA. 
From the viewpoint of the spectral shape of $\OGW^\bbh(f)$, this could imprint distortions on the power-law behaviour around the typical merging frequencies of EMRIs.
Recent analyses like \cite{Bonetti:2020jku} find that the EMRI contribution to the SWGB is still smaller than the one from stellar-mass BHs, but the uncertainty in the modelling of their formation rate is significant, and further studies will be relevant.
For what concerns the eccentricity of EMRIs, the possible presence of an AGN disk (so-called wet EMRIs) increases the friction and quickly dissipates $e$, as compared to EMRI without disk (dry EMRIs) where dynamical formation is at play and $e>0$. 
Recent studies \cite{Pan:2021oob} find the rate from wet EMRIs to dominate, thus reducing the impact of eccentric EMRIs on deviations from the $f^{2/3}$ power law.

\subsection{White dwarf binaries (BWD)}
\label{sec:BWD}
White dwarf binaries are an important class of astrophysical foregrounds for GW observations around mHz and below.
White dwarfs have a mass typically around a solar mass and are about a thousand times larger than a BH of $1\Msun$ (and accordingly less compact), so that their binary merger emits much less GW power and the peak frequency is a thousand times smaller. Although less powerful, the large number of such binaries in our own Milky way make them an important foreground for any GW experiment at $f\lesssim \mHz$. 
The foreground from extragalactic BWD mergers will be a further source of astrophysical uncertainty. It is expected to be subdominant to the background from galactic BWDs, but could be not  negligible as compared to a weakly visible primordial signal, and it is important to improve its modelling \cite{Pan:2019uyn}.

This superposition of GW signals can be controlled and reduced by identifying the loudest binary signals and removing them from the time-domain data stream.
This procedure was first exemplified by \cite{Timpano:2005gm}, to which we refer (together with more recent analyses as e.g.\,\cite{Karnesis:2021tsh,Georgousi:2022uyt,Finch:2022prg}) for a detailed discussion. 
Starting from the BWD catalogue of \cite{Nelemans:2003ha}, Ref.~\cite{Timpano:2005gm} computed the expected SGWB from their mergers. This background, from an operational point of view, adds up on top of the detector noise. It is possible to identify and remove the loudest BWDs contributing to this background: for each of the BWD of the catalogue, they computed their Signal-to-Noise Ratio (SNR) where the noise is understood as the sum of instrumental noise and BWD background. 
The loudest mergers that pass a threshold SNR are subtracted from the noise, reducing its size. 
The procedure can be iterated, and converges to an irreducible background of unresolved BWDs, after the successful removal of $\sim\mathcal O(10^4)$ of the loudest binaries.

The key feature that allows the removal of the loudest binaries is that they are approaching the merger phase (when the amplitude of the emitted GWs increases) and their GW frequency is sweeping more and more rapidly the frequency range. 
Therefore, if we consider the number of binaries from a sample population that is emitting GWs in a given frequency bin, it gets smaller as we consider bins at higher frequencies. 
When this number reduces to a few, then the corresponding BWD signals are resolvable and can be subtracted; at lower-frequency bins, the superposition of multiple signals is instead not resolvable, originating an irreducible foreground \cite{Timpano:2005gm}.
For this high-frequency bins where the expected number of sources is of order one and decreasing, the number of detected BWDs should follow the tail of a Poissonian distribution, motivating an exponential decay in the number of unresolved sources.
This is the reason why a common parametrisation \cite{Babak:2017tow, Karnesis:2021tsh,Georgousi:2022uyt,Finch:2022prg} for the SGWB from BWDs has an exponential cut-off for $f>f_\text{knee}$ (last term in this equation): 
\begin{equation}
\label{eq:BWD}
\OGW^\bwd(f) \simeq \frac{4\pi^2 f^3}{3H_0^2} \frac{A_\bwd}{\Hz} \left(\frac{f}{\Hz}\right)^{-7/3} e^{-\left(\tfrac{f}{f_1}\right)^\alpha} \left(\frac{1+\tanh\left( (f_\text{knee}-f)/f_2 \right)}{2}\right)
\end{equation}
where $H_0=h\cdot 100$ km/s/Mpc is the present Hubble rate, and $f_1$, $\alpha$, $f_\text{knee}$, $f_2$ are fit parameters.\footnote{Refs.~\cite{Cornish:2017vip, Robson:2018ifk} (with a sign correction in \cite{Schmitz:2020rag}) modify the second-to-last term in Eq.~\eqref{eq:BWD} into $\exp{\big(-(f/\Hz)^\alpha-\beta \sin(\kappa f)\big)}$, where the sin term does not have a physical interpretation and ends up being numerically almost irrelevant for the best-fit parameters. 
Also, we consider physically more appropriate to include a free-parameter $f_1\sim \mHz$ in the fit, rather than fixing it to 1 Hz where $\OGW^\bwd$ is totally negligible.
We decide then to follow the more recent convention in Eq.~\eqref{eq:BWD}. 
At any rate, this factor has the only purpose of accommodating a better fit to the residual confusion noise around the knee frequency.}
The knee-frequency where the confusion noise drops slowly decreases with the time duration of the mission, as more BWDs can be identified and subtracted. 
For our analysis, we adopt from \cite{Georgousi:2022uyt} (which is based on the BWD catalogue from \cite{Korol:2021pun}) the best-fit values $A_\bwd=2.7\cdot 10^{-44}$, $f_1=0.64\mHz$, $\alpha=1.26$, $f_\text{knee}=2.0\mHz$, $f_2=0.28\mHz$.

In summary, for our purposes the relevant point about the confusion noise of galactic BWDs is that, after removal of the resolvable loudest binaries, the residual foreground falls exponentially at $f\sim \mHz$, around the peak sensitivity of LISA. 
It is therefore safe to assume that the frequency range above this value will not be significantly contaminated by the foreground from BWDs \cite{Pan:2019uyn}.
The exponential drop in $\OGW^\bwd$ is the main spectral feature that is relevant in the following.

\section{Primordial backgrounds of gravitational waves for precision cosmology}
\label{sec:prim}

Among the possible primordial GW backgrounds that could populate our Universe, which include backgrounds generated during inflation
\cite{Starobinsky:1979ty, Rubakov:1982df, Grishchuk:1993te,Turner:1996ck, Bartolo:2016ami, Guzzetti:2016mkm},
the subsequent reheating phase~%
\cite{Khlebnikov:1997di,Easther:2006gt,Easther:2006vd,GarciaBellido:2007dg,GarciaBellido:2007af,Dufaux:2007pt}%
, phase transitions (PT)~%
\cite{Caprini:2019egz} %
and by topological defects (such as cosmic strings)~%
\cite{Caprini:2018mtu}%
, or finally by scalar perturbations at 2nd order in perturbation theory~%
\cite{Acquaviva:2002ud,Mollerach:2003nq,Baumann:2007zm,Espinosa:2018eve,Kohri:2018awv,Domenech:2019quo}, in this paper we are interested in the exciting prospect of precision measurements of cosmological parameters from the GW spectrum.
Many studies discussed related effects and their observability in various scenarios~\cite{%
Seto:2003kc,
Boyle:2005se,
Watanabe:2006qe,
Boyle:2007zx,
Jinno:2012xb,
Caprini:2015zlo,
Geller:2018mwu,
Saikawa:2018rcs,
Cui:2018rwi,
Caldwell:2018giq,
DEramo:2019tit,
Figueroa:2019paj,
Auclair:2019wcv,
Chang:2019mza,
Caprini:2019egz,
Gouttenoire:2019kij,
Gouttenoire:2019rtn,
Allahverdi:2020bys,
Blasi:2020wpy,
Domenech:2020kqm%
}.
Similar effects, like variations in the number of relativistic degrees of freedom $g_\star$, or in the equation-of-state parameter $w$ of the universe, manifest themselves as modifications to the spectral shape of a primordial SGWB.
To this end, we should know the primordial shape of the signal, before the cosmological history affected it. 
This can be confidently assumed for the class of GW backgrounds generated by phenomena that occur locally in space and time, independently in each Hubble patch.
In this regime, the principle of causality fixes the spectral shape of the SGWB at low frequencies, corresponding to wavelengths larger than the correlation length of the GW source \cite{Hook:2020phx}.
In a radiation-dominated Universe, and if the GW propagate as free waves, the spectrum at low frequencies must go as $f^3$ \cite{Caprini:2009fx}. 
As illustrated in \cite{Hook:2020phx, Brzeminski:2022haa}, this behaviour can only be affected by the evolution of the universe (i.e.~by modifications to $w$, or $g_\star$ and hence to the Hubble rate) or by the propagation of the GWs, which can be damped by anisotropic stress sourced e.g.~by free-streaming relativistic particles.
The primary example of a causality-limited source of a SGWB is a first-order PT.
In this paper, we consider for concreteness a primordial GW signal sourced by a PT, and we assess the sensitivity to cosmological parameters from the causality-limited tail of the spectrum, similarly to \cite{Brzeminski:2022haa} but accounting also for the presence of astrophysical GW foregrounds. 

We consider a PT occurring when the Universe has a temperature $\Ts$, whose duration is $\beta^{-1}$ and releasing an energy fraction $\alpha$ of the total energy density of the Universe.
The GW energy spectrum consists of the contributions from various sources (bubble collisions, sound waves and turbulence in the plasma; for recent reviews, see \cite{Caprini:2019egz, Hindmarsh:2020hop}); for simplicity, we consider the contribution that is typically the largest, i.e.~from sound waves in the plasma.
The results of numerical simulations can be approximated by the analytical formula \cite{Hindmarsh:2017gnf}
\begin{equation}
\label{eq:prim}
h^2 \OGW^\pt(f) = 1.2 \cdot 10^{-6} 
  \left( \frac{\Hs}{\beta}\right) \left( \frac{\kappa_v \alpha}{1+\alpha}\right)^2 \left( \frac{106.75}{g_\star}\right)^{\frac 13} 
  \left( \frac{f}{f_*}\right)^3   \left( \frac{7}{4+3(f/f_*)^2}\right)^\frac 72
\cdot \mathcal F(f,\theta)
\end{equation}
where the peak frequency $f_*$ is roughly of the order of the Hubble rate at $\Ts$, and corresponds to a present-day frequency
\begin{equation}
\label{eq:f*}
f_* = 9 \cdot 10^{-3}\mHz \frac{1}{v_w} \left( \frac{\beta}{\Hs}\right) \left( \frac{\Ts}{100 \GeV}\right) \left( \frac{g_\star}{106.75}\right)^{\frac 16} \,.
\end{equation}
The energy fraction $\alpha$ mostly affects the normalisation of $\OGW^\pt$, and the inverse duration of the PT $\beta$ shifts $\OGW^\pt$ both vertically and horizontally.
Other parameters of the PT are the latent heat fraction that gets converted into bulk motion of the fluid $\kappa_v$, and the bubble wall terminal velocity $v_w$. 
For definiteness, we fix the following values for the PT parameters:
\begin{equation}
\beta= 4 \Hs\,, \quad
\alpha=0.1 \,, \quad
v_w=1\,, \quad
g_\star=106.75 \,, \quad
\kappa_v = \frac{\alpha}{0.73+0.083\sqrt \alpha + \alpha} \,.
\end{equation}

We consider the following possible modifications to the low-frequency tail of the GW spectrum.
In the case when the equation-of-state parameter of the Universe is $w\neq \tfrac 13$ at the time when the Hubble rate is equal to a frequency $f_w \leq f_*$ shortly after the PT, the prefactor is \cite{Hook:2020phx}%
\footnote{Strictly speaking, also the physical frequency $f$ is affected if we modify $w$ with respect to the radiation-dominated case, as the total expansion factor of the universe changes. Given that such an effect is small, and that most of its impact is degenerate to a change in the unknown temperature $\Ts$, we neglect it \cite{Brzeminski:2022haa}.}
\begin{equation}
\label{eq:prim w}
\mathcal F(f, w) = \left(\frac{f}{f_w}\right)^2 \left(
  \left|j_\frac{1-3w}{1+3w}\left(\frac{f}{f_w}\right)\right|^2 +
  \left|y_\frac{1-3w}{1+3w}\left(\frac{f}{f_w}\right)\right|^2 
  \right)
\end{equation}
where $j_n(z)$, $y_n(z)$ are spherical Bessel functions (notice that $\mathcal F(f,w=\tfrac 13)=1$). 
The impact of this factor is shown in Fig.~\ref{fig:PT-w} for a few values of $w$: for a standard radiation-dominated cosmology, $w=\tfrac 13$ and the scaling of the SGWB at low frequencies is $f^3$, while for different values of $w$ the spectral tilt is $\frac{1+15w}{1+3w}$.

Another unique probe of the early Universe that is offered by the causality-limited tail of the GW spectrum is the measurement of the fraction $\fFS$ of energy density of relativistic free-streaming particles around $\Ts$. 
In presence of free-streaming particles that modify the propagation of GWs in the universe, the modification to the primordial background in Eq.~\eqref{eq:prim} is
\begin{equation}
\label{eq:prim fFS}
\mathcal F(f, \fFS) = \left(\frac{f}{f_*}\right)^2 \left(
  \left|j_{\alpha_\textsc{fs}}\left(\frac{f}{f_*}\right)\right|^2 +
  \left|y_{\alpha_\textsc{fs}}\left(\frac{f}{f_*}\right)\right|^2 
  \right) \,, \quad \alpha_\textsc{fs} = -\frac 12 + \sqrt{\frac 14 - \frac 85 \fFS}
\end{equation}
where $f_*$ is the peak frequency given in Eq.~\eqref{eq:f*}. 
In the absence of free-streaming particles, $\mathcal F(f,\fFS=0)=1$, whereas for $\fFS>0$ the tilt at low frequencies becomes steeper, from $3$ to $3+\tfrac{16}{5} \fFS$ (and for $\fFS>\tfrac{5}{32}$ saturates to $4$, with an additional small oscillating pattern imprinted on the power-law spectrum \cite{Hook:2020phx})%
\footnote{If the free-streaming particles become non-relativistic later on, the GW frequencies entering the horizon afterwards are not significantly impacted by the effect of $\fFS$, and the GW spectrum resumes the $f^3$ behaviour at low frequencies \cite{Brzeminski:2022haa}.}.

\section{Optimal estimator for a primordial SGWB in presence of astrophysical foregrounds}
\label{sec:method}
As we discussed in Sec.~\ref{sec:astro fg}, the two main astrophysical foregrounds in the frequency range $\mHz-\Hz$ are the unresolvable superpositions of GWs from mergers of BBHs and BWDs, and we justified why both their spectral shapes have two distinguishing features.
The BBH background can be approximated to a good accuracy as a $f^{2/3}$ power law down to $\OGW^\bbh(f)\gtrsim 10^{-12}$ where small deviations from $\OGW^\bbhe(f)$ can be expected (see Eq.~\eqref{eq:BBH ecc}).
The BWD background (arising from the confusion noise of binaries that cannot be isolated in a frequency bin and subtracted), is supposed to fall off exponentially around $f_\text{knee}$ (see Eq.~\eqref{eq:BWD}).
Therefore, we argue that the leading uncertainties in these astrophysical foregrounds are encapsulated in one free parameter each, that is their overall normalisation. 
This will be our working assumption in the remainder of the paper, and we have provided justifications for this hypothesis in the discussion of Sec.~\ref{sec:astro fg}.
We then assess the sensitivity to a primordial signal by performing a simple Fisher matrix analysis.
The optimal filter, that enhances the SNR for a given signal by accounting for the presence of astrophysical foregrounds, is derived in \cite{Poletti:2021ytu}.
We summarise here those results and we refer to \cite{Poletti:2021ytu} for a complete and detailed discussion. 

Let us denote the signal $d_I(t)$ measured in the channel $I$ of a GW detector ($I$ can run e.g.\,over the two independent channels of the LISA proposal, or over the interferometers of LVK) as the sum of a GW signal $s_I(t)$ and the noise $n_I(t)$ in the detector:
\begin{equation}
d_I(t) = s_I(t) + n_I(t) \,,
\end{equation}
where the the signal $s_I(t)$ is related to the GW in frequency space $h_P(f,\hat n)$ ($P$ being the polarisation $+,\times$, and $\hat n$ the unit vector in the direction of propagation $\vec k$) through 
\begin{equation}
\label{eq:signal s}
s_I(t) = \sum_{P=+,\times} \int_{-\infty}^{+\infty} \d f \int \d^2\hat n \, F_I^{(P)}(\hat n, f) \,h_P(f,\hat n)\, e^{i 2\pi f (t-\hat n\cdot \vec x_I)} \,,
\end{equation}
where the response function $F_I^{(P)}(\hat n, f)$ of a channel $I$ depends on the properties of the detector.
We define the spectral densities of signal and noise,
\begin{align}
\big\langle h^*_P(f,\hat n) h_{P'}(f,\hat n') \big\rangle &= \tfrac{1}{8\pi} \delta^{(2)}(\hat n-\hat n') \,\delta_{P,P'}\, \delta(f-f')\, \Sh(f) \,,\\
\label{eq:N}
\big\langle n_I(f) n_J^*(f') \big\rangle &= \tfrac{1}{2} \, \delta(f-f')\, N_{IJ}(f) \,.
\end{align}
The signal can be extracted from the correlation estimator
\begin{equation}
x_{IJ} = \int_{-T/2}^{T/2} \d t \int_{-T/2}^{T/2} \d t'
    \Big(d_I(t) d_J(t') -\big\langle n_I(t) n_J(t')\big\rangle \Big) Q_{IJ}(t,t')\,,
\end{equation}
where $T$ is the time duration of the experiment (we take $T=4y$ for LISA), and $Q$ is an arbitrary filter function that we choose in  order to maximise the sensitivity to signal.
We now define the signal spectral density as the sum of a primordial and an astrophysical component (we discuss here for simplicity the case of one astrophysical foreground, and we refer to App.~A in \cite{Poletti:2021ytu} for the formul\ae\ in the case of multiple foregrounds)
\begin{equation}
\Sh(f)=\Sp(f)+ \Aa \Sa(f) \,,
\end{equation}
where $\Aa$ is an unknown normalisation parameter for which some information is available. 
In particular, we denote by $\avAa$ its expected value and by $\sa^2$ its variance.
We can introduce a modified estimator in presence of astrophysical foreground
\begin{equation}
y_{IJ} = x_{IJ} - \avAa T \int_0^\infty \d f \Sa(f)\, \R_{IJ}(f)\, Q_{IJ}(f)\,,
\end{equation}
where $\R$ is the response function defined  by
\begin{equation}
\label{eq:response function}
\mathcal R_{IJ}(f) \equiv 
\frac{1}{4\pi} \int \d^2\hat n
\sum_{P=+,\times}  F_I^{(P)*}(\hat n, f) \,F_J^{(P)}(\hat n, f)
e^{-i 2\pi f \,\hat n \cdot(\vec x_I- \vec x_J)}\,,
\end{equation}
depending on the properties of the detectors $I,J$ and their relative orientation and distance.
The expectation value of $y_{IJ}$ gives the primordial signal in frequency space:
\begin{equation}
\langle y_{IJ}\rangle = T \int_0^\infty \d f \Sp(f) \, \R_{IJ}(f)\, Q_{IJ}(f) \,.
\end{equation}
We now define the signal-to-noise ratio SNR as
\begin{equation}
\label{eq:SNR}
\SNR  = \frac{\langle y_{IJ} \rangle}{\sqrt{\langle y_{IJ}^2 \rangle -\langle y_{IJ} \rangle^2}} \,,
\end{equation}
and we want to choose an optimal filter function $Q_{IJ}(f)$ in frequency space to maximise the SNR for a given primordial signal.
It is possible to show that this is equal to the Fisher information on the amplitude of the signal, under the assumption of Gaussianity \cite{Romano:2016dpx,Caprini:2019pxz,Flauger:2020qyi}.

We can express the final result of \cite{Poletti:2021ytu} in the following compact form, which allows a geometrical interpretation of the result.
First, we define the following scalar product between integrable functions in frequency space,
\begin{equation}
\label{eq:scalar product}
\bm{u} \cdot \bm{v} \equiv 2T \int_0^\infty \d f \, \N_{IJ}^2(f) \,u^*(f) \,v(f) \,,
\end{equation}
where the noise spectral density $\N_{IJ}^2(f)$ is given by
\begin{equation}
\label{eq:noise spectral density}
\N_{IJ}^2(f) \equiv N_{II} (f) \, N_{JJ}(f) + N_{IJ}^2 (f)\,.
\end{equation}
We also denote
\begin{equation}
\s \equiv \frac{\R_{IJ}(f)}{\N_{IJ}^2(f)}\Sp(f) \,, \quad 
\a \equiv \frac{\R_{IJ}(f)}{\N_{IJ}^2(f)}\Sa(f) \,.
\end{equation}
We can then write the optimal filter function in presence of astrophysical foregrounds simply as 
\begin{equation}
\label{eq:Q astro}
Q_{IJ}(f) = \left( \s - \a \frac{\s\cdot \a}{\sa^{-2}+\a\cdot \a} \right) \,,
\end{equation}
as opposed to the standard optimal filter in the absence of foregrounds, which is just $Q_{IJ}(f)=\s$.
The optimal filter ranges from $\s$ in the limit $\sa\to 0$ (perfectly known foreground) to a function becoming increasingly orthogonal to $\a$ for larger $\sa$.
In other words, in presence of foregrounds, the optimal filter counterweights the astrophysical components by subtracting a component proportional to $\a$ while computing the SNR.
This can be nicely seen in the illustrative example of \cite[Fig.~3]{Poletti:2021ytu} where, for growing values of $\sa$, $Q_{IJ}(f)$ becomes negative across some frequency ranges, so that the filter does not pick up power from foreground-like spectral shapes.

Finally, the SNR in presence of an astrophysical foreground can be simply written as
\begin{equation}
\label{eq:SNR astro}
\SNR^2 = \s \cdot \s - \frac{(\s \cdot \a)^2}{\sa^{-2}+\a \cdot \a} \,,
\end{equation}
which highlights the reduction of the SNR when the foreground $\a$ has a ``parallel'' component to the signal $\s$, i.e.\,when the spectral shapes of signal and foreground are similar.
Two limits are instructive to consider. 
The SNR approaches its value $\s\cdot\s$ without foregrounds when $\s\cdot \a \ll \s\cdot \s$, that is when $\s$ and $\a$ are dominated by different frequency ranges and their shapes do not significantly overlap.
Another limit in which the SNR recovers the value without foregrounds is when $\sa\to 0$: in that case, the foreground is perfectly known and our sensitivity to any signal sitting above it is only mildly affected.

We are interested in the case in which the unknown parameter $\theta$ that we want to measure in the signal is not the amplitude of such signal (that we assume to be measured), but a cosmological parameter $\theta =w$ or $\fFS$ distorting the low-frequency tail of the signal (see Sec.~\ref{sec:prim}), that affects the signal non-linearly as in Eqs.~\eqref{eq:prim w} and \eqref{eq:prim fFS}.
We can apply the previous results to the estimate of the sensitivity to the parameter $\theta$ by including $\s$ among the templates (in the generalisation of Eq.~\eqref{eq:SNR astro} to multi-component subtraction, see \cite{Poletti:2021ytu}), and inserting  $\partial_\theta \s$ (rather than $\s$) in Eq.~\eqref{eq:SNR astro}. Under the same assumptions mentioned above, the $\SNR^2=1/\sigma_\theta^2$ can be shown to be the Fisher information on $\theta$.
We have now introduced all the ingredients to compute the sensitivity to cosmological parameters in a primordial GW signal, in presence of astrophysical foregrounds.

\section{Results}
\label{sec:results}
We now apply the procedure illustrated in detail in \cite{Poletti:2021ytu} and outlined in Sec.~\ref{sec:method} to the scenario that we have been considering throughout this paper.
We would like to quantify the sensitivity to the spectral features in the low-frequency (causality-limited) range of the primordial SGWB generated by a local phenomenon like a PT, in presence of astrophysical GW foregrounds.

We have discussed in Sec.~\ref{sec:astro fg} why it is a good assumption to treat the SGWB spectra from BWD and from BBH mergers as known up to a normalisation factor, and up to a contribution from eccentric BBHs that we have quantified in Eq.~\eqref{eq:BBH ecc} and that we separately keep into account as $\OGW^\bbhe$ in the following.
Therefore we model $\OGW^\bbh(f)$ and $\OGW^\bwd(f)$ as given by Eqs.~\eqref{eq:bbh norm},~\eqref{eq:OGW 2/3} and \eqref{eq:BWD} up to the normalisation factors not being perfectly known, and we assume them to lie within fractional uncertainties $\sigma_\bbh$, $\sigma_\bwd$ from their central values.

For what concerns the contribution $\OGW^\bbhe$ from eccentric BBH mergers, we model it as a constant foreground in frequency space, with a normalisation given in Eq.~\eqref{eq:BBH ecc} with an uncertainty $\sigma_\bbhe$. 
This assumption is partly motivated by expectation that, in the frequency range of LISA, its spectrum should not vary by orders of magnitude, and partly by the results of \cite{Perigois:2021ovr}%
\footnote{We thank the authors of \cite{Perigois:2021ovr} for providing us with the data corresponding to their Fig.~4 to compute $\OGW^\bbhe(f)$.}%
, which find a featureless spectrum. 
We do not expect this assumption to impact much on our results. The reason is that, as long as the actual component $\OGW^\bbh(f)$ is not similar in spectrum to the primordial signal, and is limited from above not to exceed a value of the order of Eq.~\eqref{eq:BBH ecc}, its precise shape does not modify significantly the optimal filter function. 
This statement can be reinforced in the future by means of further progress from the joint effort of astrophysical modelling and GW observations to corner down the uncertainties on the contribution from eccentric binaries.

\begin{figure}[h!]\centering
\includegraphics[width=0.7\textwidth]{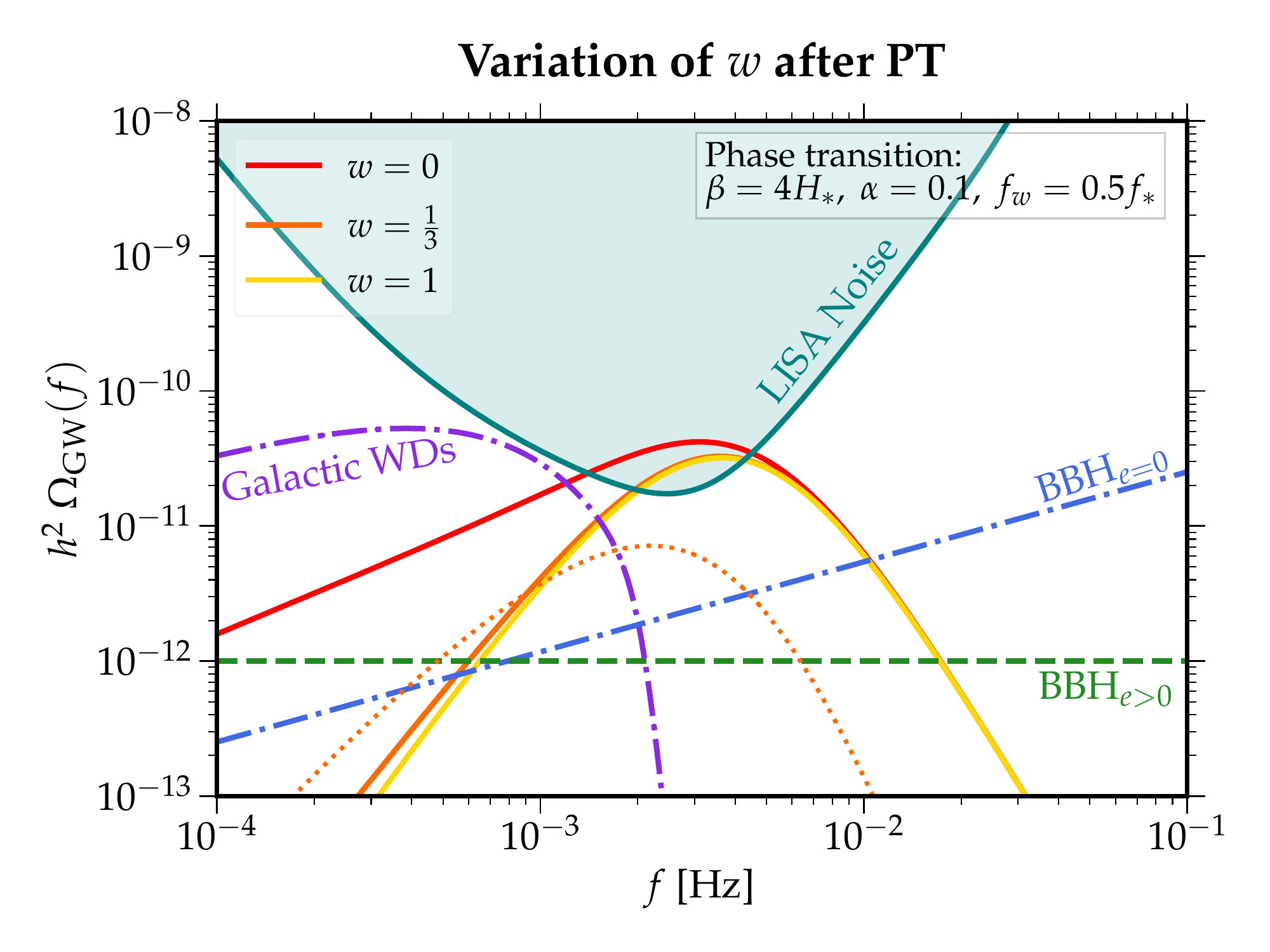}\vspace{-1em}
\hspace*{-10pt}
\includegraphics[width=0.53\textwidth]{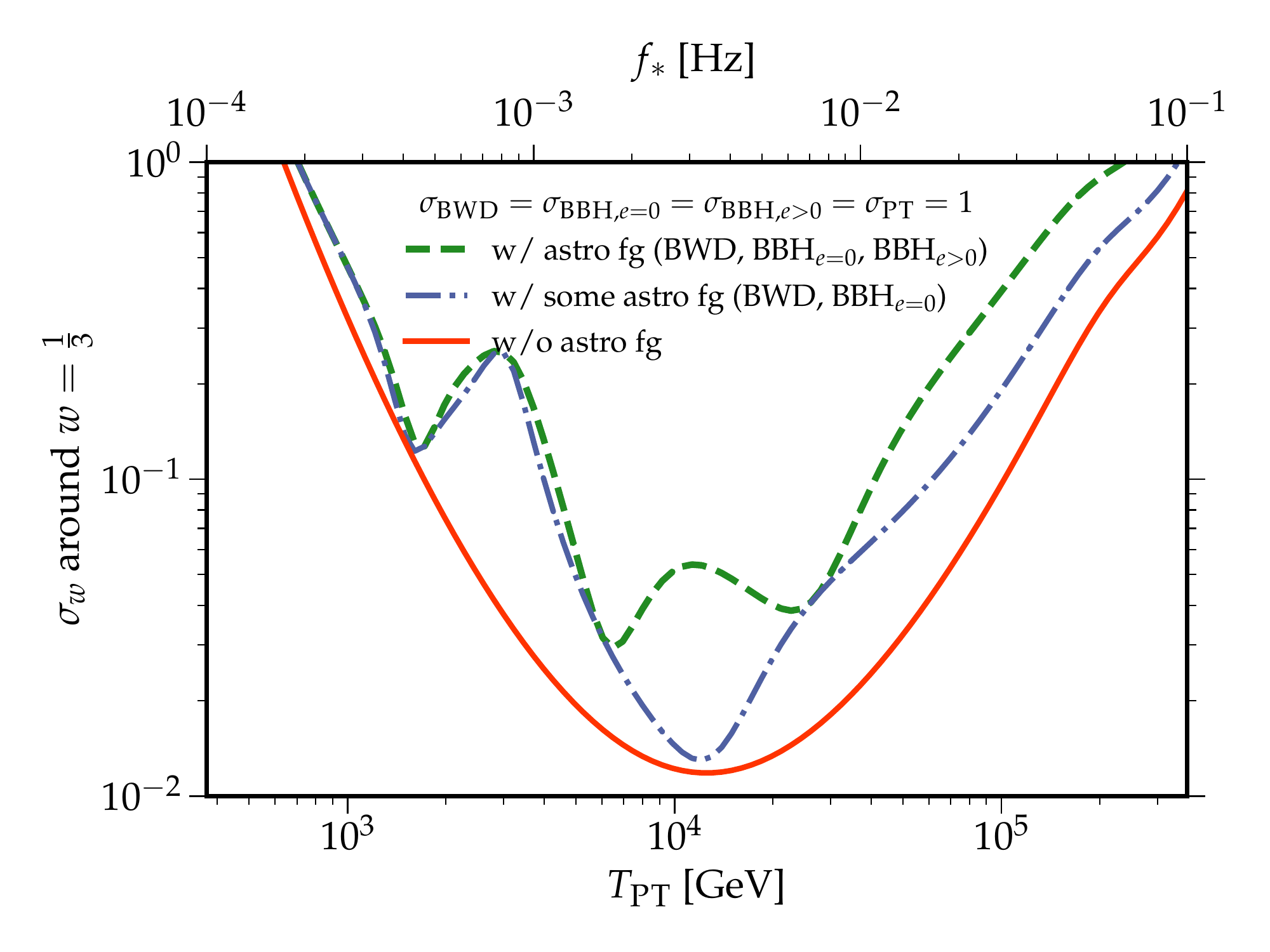}
\hspace*{-50pt}\hfill
\includegraphics[width=0.53\textwidth]{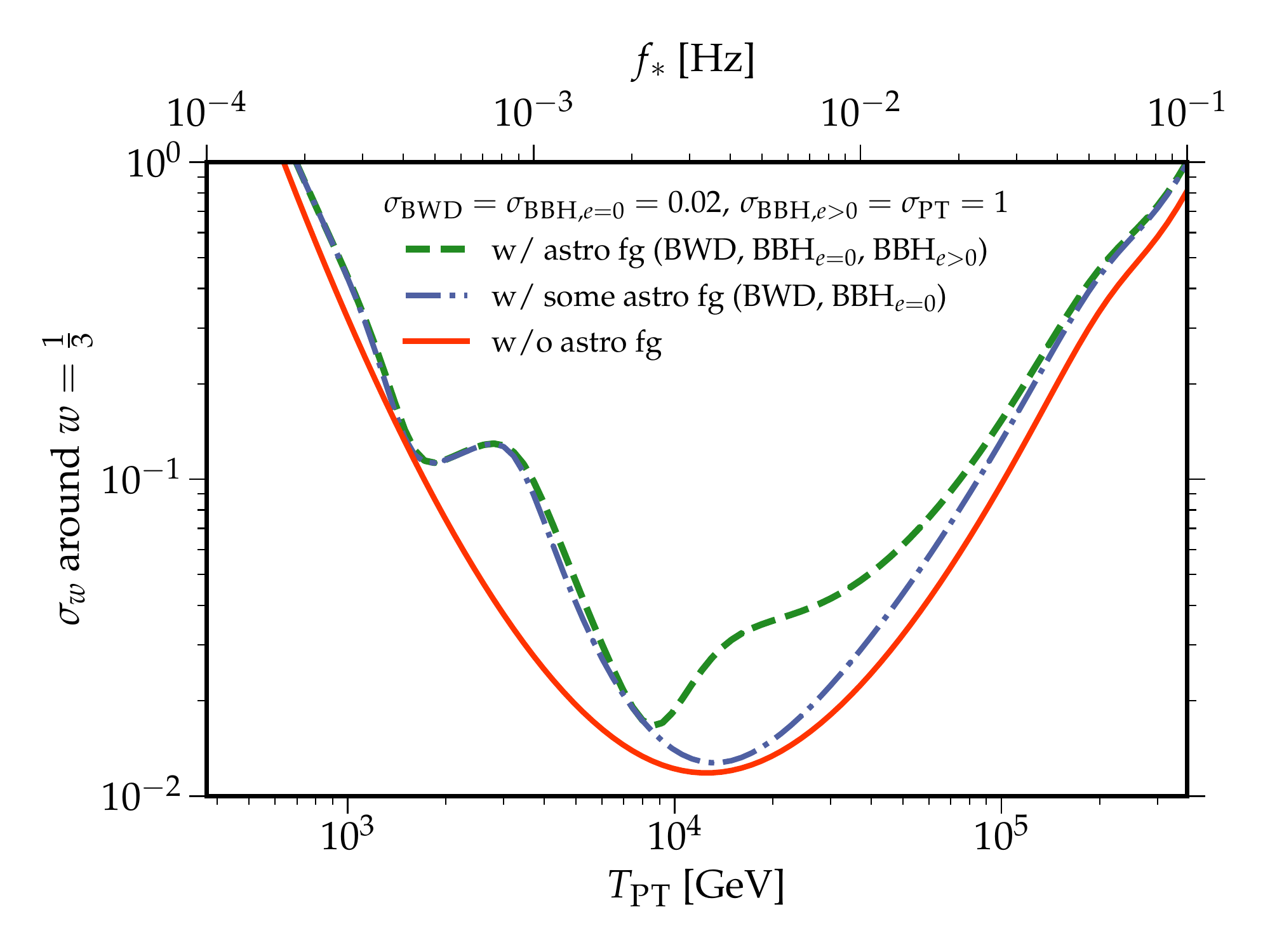}
\hspace{-10pt}
\caption{Impact of the uncertainty of astrophysical foregrounds on the measurement of a primordial GW background signal.
\textit{Top plot}: comparison of the noise sensitivity of LISA to $\OGW(f)$ (green shaded curve), main astrophysical foregrounds in this frequency range (binary White Dwarfs in purple, Binary Black Holes extrapolating $\OGW^\bbh \propto f^{2/3}$ in blue, and contribution from eccentric binaries in green; see Sec.~\ref{sec:astro fg}), a primordial Phase transition followed by a phase with $w\neq\tfrac 13$ (red-orange curves, and $\partial_w \OGW^\pt\big|_{w=1/3}$ is shown by a dotted line; see Sec.~\ref{sec:prim}).
\textit{Bottom left plot}: sensitivity to $\sigma_w$ around $w=\tfrac 13$ without including the uncertainty on astrophysical foregrounds (red curve), accounting for BWDs and BBHs (blue dot-dashed), and by also adding the background from eccentric BBHs (green dashed).
The relative uncertainties on the astrophysical foregrounds are taken to be 1; their exact value does not matter, as the optimal filter removes all the signal component parallel to the foreground (see Eq.~\eqref{eq:SNR astro}).
\textit{Bottom right plot}: the same as the  left plot, but with $\sigma_\bbh=\sigma_\bwd =2 \%$, to show the typical fractional uncertainty below which astrophysical foregrounds are known well enough not to degrade the sensitivity to the signal.
}
\label{fig:PT-w}
\end{figure}

\begin{figure}[h!]\centering
\includegraphics[width=0.7\textwidth]{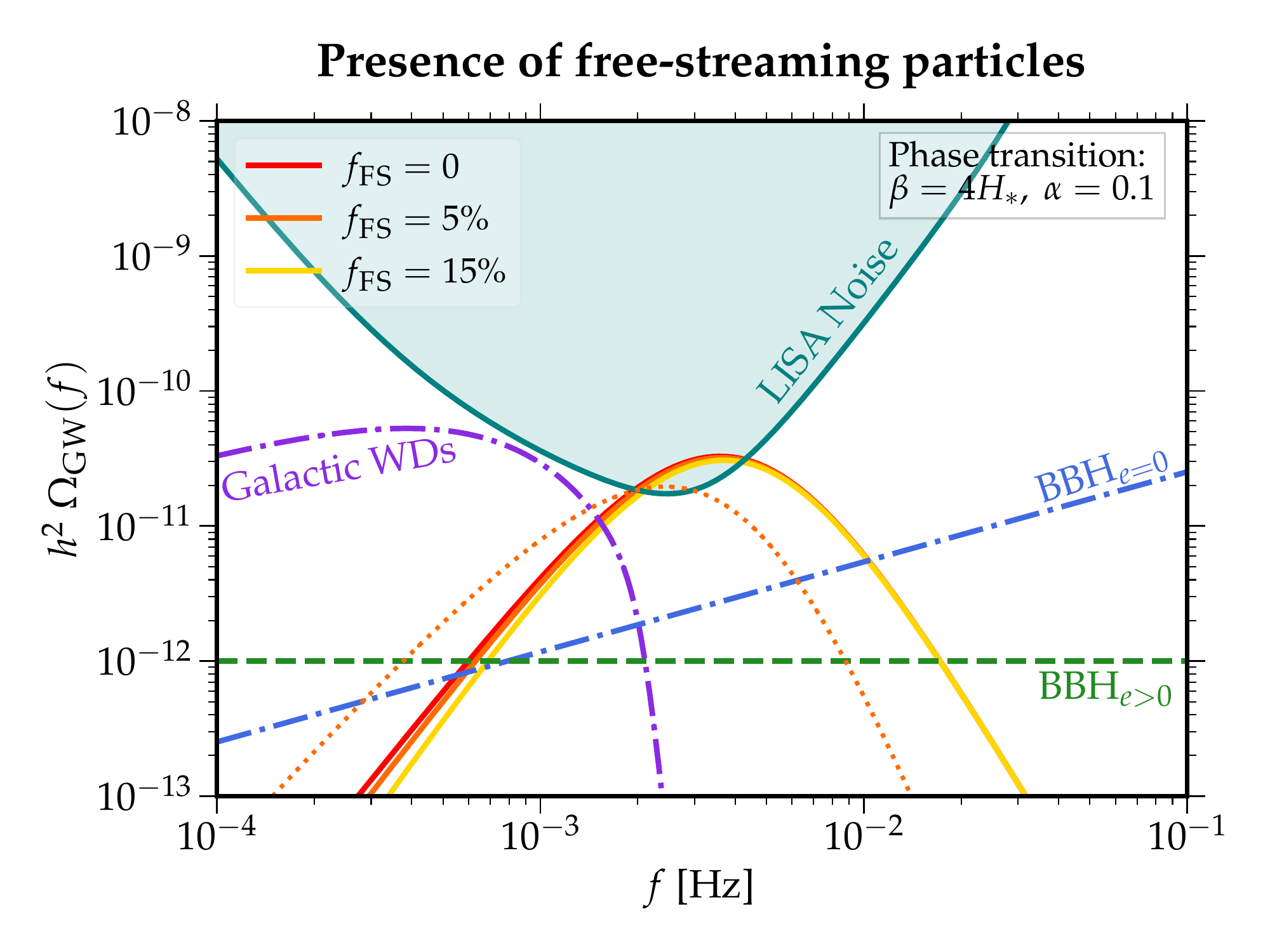}\vspace{-1em}
\hspace*{-10pt}
\includegraphics[width=0.53\textwidth]{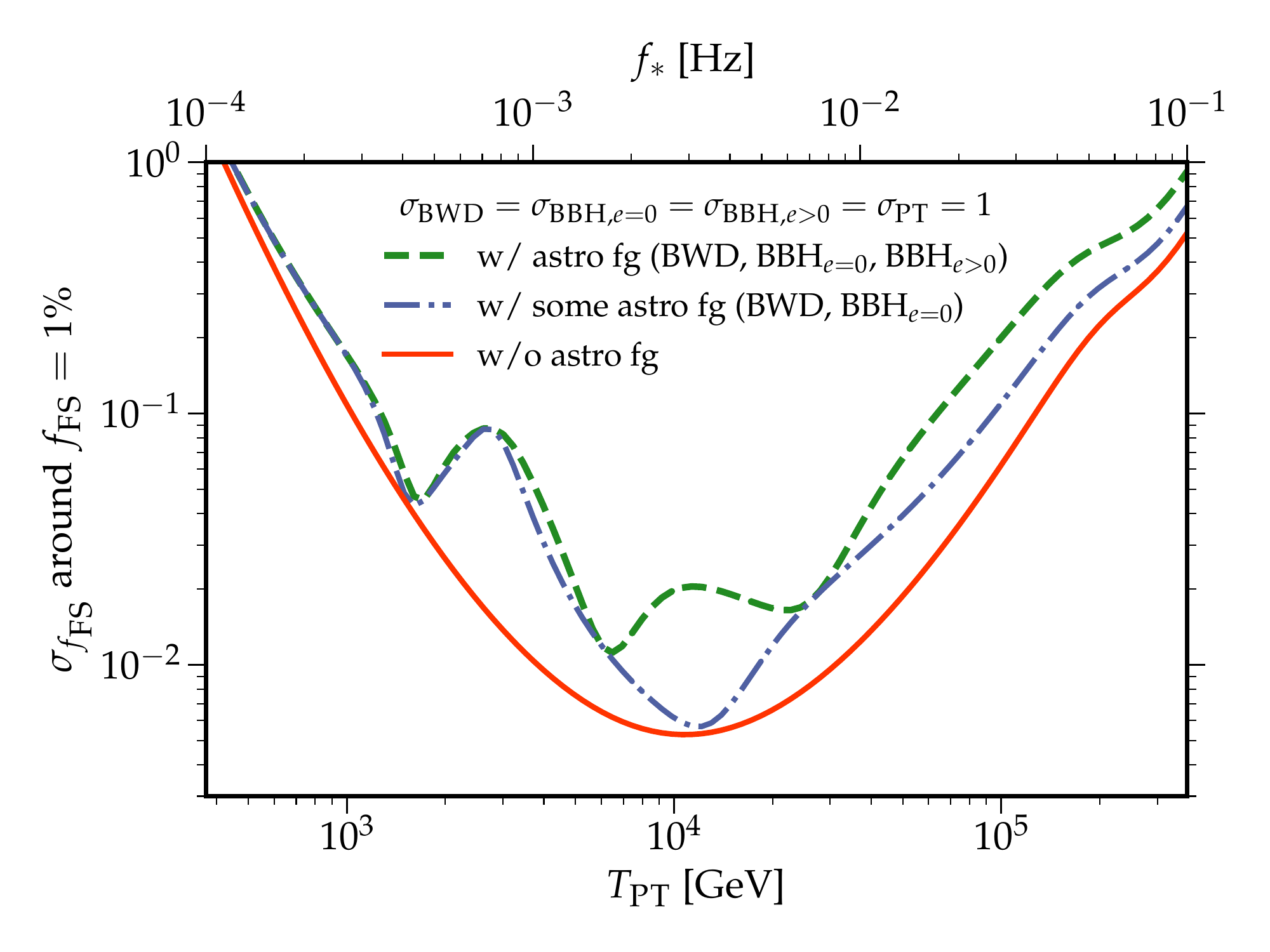}
\hspace*{-30pt}\hfill
\includegraphics[width=0.53\textwidth]{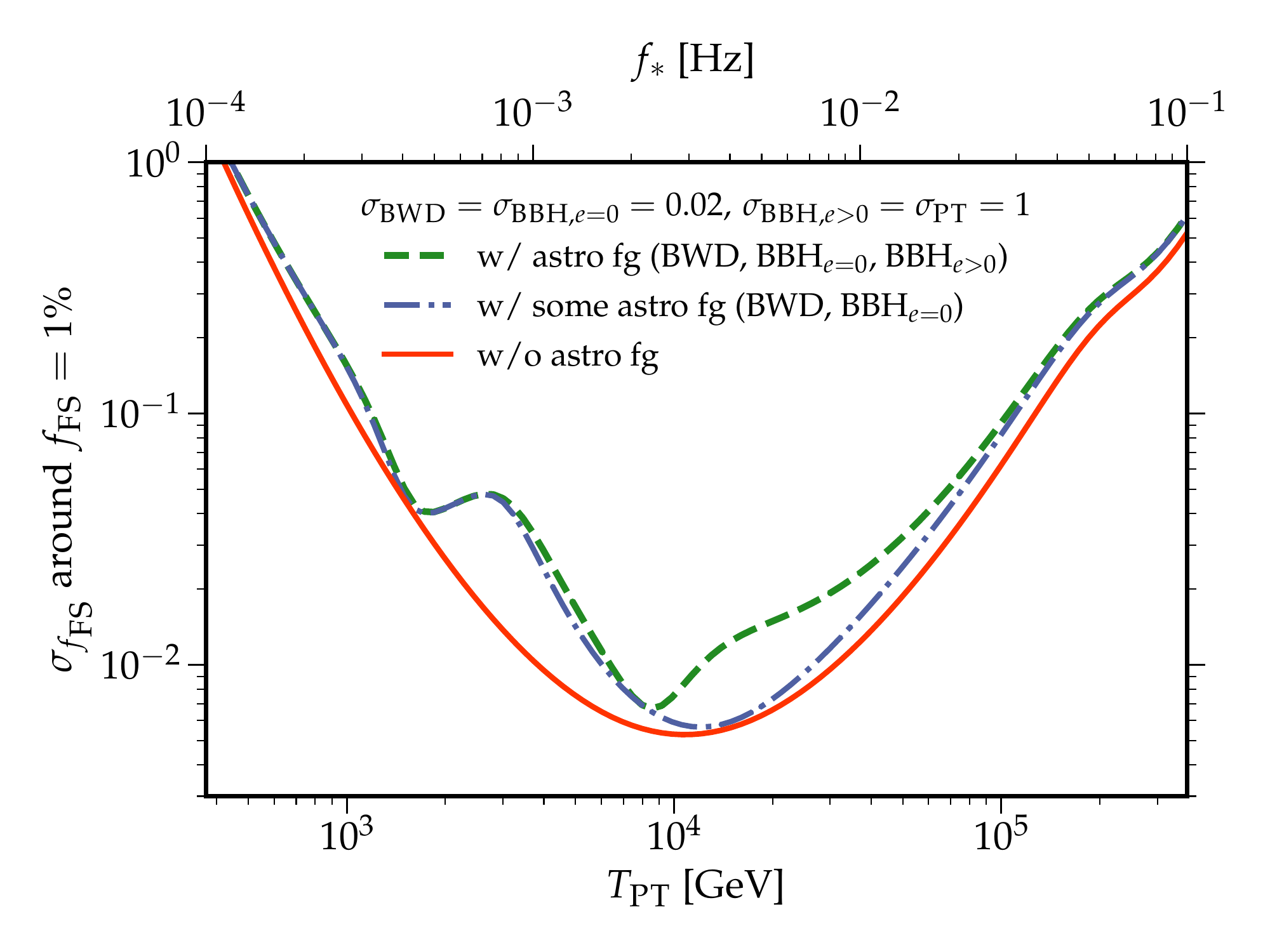}
\hspace{-10pt}
\caption{The same as Fig.~\ref{fig:PT-fFS}, but looking for the presence of an energy fraction $\fFS$ of free-streaming particles in the Universe at the time of the phase transition. The top plot shows three reference values for $\fFS$, and we consider in the bottom plots the sensitivity to $\sigma_{\fFS}$ around $\fFS=1\%$.
}
\label{fig:PT-fFS}
\end{figure}

We show the results of our analysis for two significant cases of primordial signals, as we discussed in Sec.~\ref{sec:prim}.
We consider the potential to perform precision cosmology after the discovery of a primordial signal from a PT, for which we assume a spectral shape as in Eq.~\eqref{eq:prim}.
Two key observables are the equation-of-state parameter $w$ of the Universe for a few $e$-folds after the PT, and the energy fraction $\fFS$ of relativistic free-streaming particles in the Universe at the time of the PT.
The two cases are shown respectively in Figs.~\ref{fig:PT-w} and \ref{fig:PT-fFS}. 

In each figure, the \textit{top panel} shows:
\begin{itemize}
\item the LISA noise sensitivity curve (for the AA and EE channels; we refer to \cite{Smith:2019wny} for the calculation of LISA noise curves and response functions). 
We remind the reader that these strain sensitivity curves should not be confused with the power-law integrated sensitivity curves (that stretch down to $\OGW \sim 10^{-13}$ for LISA, see e.g.~\cite[Fig.~2]{Caprini:2019pxz}), which represent the envelope of the upper limits at each frequency on the GW power-law signal that is detectable with a given SNR. While the latter allows for a quick graphical check about whether a GW background is detectable, the original noise sensitivity curve is required to compute the SNR;
\item the primordial signal for a few values of $w$, $\fFS$ and for the PT parameters listed in the box on the upper right. The dotted curve, for reference, shows the input for the calculation of the sensitivity to $\theta=w,\fFS$, that is $\partial_\theta \OGW^\pt(f)$ (see Sec.~\ref{sec:method}); 
\item the three astrophysical foregrounds discussed in Sec.~\ref{sec:astro fg} (BWDs, BBHs, eccentric BBHs).
\end{itemize}
The lower panels of Figs.~\ref{fig:PT-w} and \ref{fig:PT-fFS} show the main results of this paper, that is the sensitivity to variations in $w$ (or $\fFS$) around a reference value.
For simplicity, in these plots we vary only this parameter, and keep the other PT parameters fixed: the reason is that the impact of $w,\fFS$ (which affects the low-frequency range of the spectrum) is quite distinct from the other parameters, which do not alter the causal part of the primordial SGWB, so there aren't significant degeneracies.
The three curves show $\sigma_w$ (or $\sigma_{\fFS}$) for the following assumptions:
\begin{itemize}
\item red curve: without accounting for the uncertainty on any astrophysical foreground (the results are in agreement with \cite{Brzeminski:2022haa}). This curve is peaked for $f_*$ around the maximum sensitivity of LISA (around a few mHz) where the signal is highest above the noise, and is weaker elsewhere, roughly mirroring the noise curve of LISA;
\item blue dot-dashed curve: accounting for the unknown normalisation of $\OGW^\bwd(f)$ and $\OGW^\bbh(f)$ with the fractional uncertainties $\sigma_\bwd$, $\sigma_\bbh$ listed in the legends.
The bottom left plot assumes these astrophysical uncertainties to be $\sim \mathcal O(1)$: their actual value does not matter quantitatively, because for this size of $\sigma$, the component of the signal parallel to the foregrounds is discarded entirely (see Eq.~\eqref{eq:SNR astro}), which makes the forecast largely insensitive to the actual amplitude of the foregrounds.
The impact of a foreground on the deterioration of the SNR depends on its degeneracy in frequency shape with the primordial signal.
The bottom right plot shows the values of $\sigma_\bwd$, $\sigma_\bbh$ (below $\sim 5\%$ of fractional uncertainty) at which the astrophysical foregrounds are constrained enough that the degradation of the SNR starts to be less significant.
\item green dashed curve: the same as the previous curve, with the inclusion of $\OGW^\bbhe$ from the eccentric binaries. 
We keep $\sigma_\bbhe=1$ in both the bottom plots, as this background suffers from larger uncertainties.
\end{itemize}

We can draw the following conclusions from the results of Figs.~\ref{fig:PT-w} and \ref{fig:PT-fFS}.
As we can see from the bottom right plots, in order to accurately detect a peaked primordial signal we need to know the normalisation of the astrophysical signal with an accuracy around a few percent, for its impact on the SNR to be negligible.
Such a precision is unrealistic, given the wide set of astrophysical uncertainties that are encapsulated in the normalisation of these foregrounds, so we cannot neglect these uncertainties and they have an impact on our analysis.
As we quantitatively assess their impact, though, we can see from the bottom left plots that, even when their normalisation is highly uncertain, the knowledge of their spectral shape strongly reduces the degradation of the SNR. The increase in $\sigma_w$, $\sigma_{\fFS}$ is by a factor of a few, and only matters where the primordial and astrophysical signals peak around the same frequencies.
As long as our modelling of the astrophysical foregrounds allows us to exclude ``non-standard'' contributions with a shape (and amplitude) comparable to the primordial signal we are looking for, those uncertainties do not hinder our discovery potential with GW detectors.

\section{Conclusions}
\label{sec:conclusions}
As we advance in the era of GW astronomy, ongoing measurements of the BBH merger rate from LVK improve the estimate of the expected GW background from unresolved BBH mergers, that should be eventually measured by the collaboration at the achievement of their design sensitivity.
This measurement will corner down part of the uncertainties on the astrophysical foreground from BBH mergers that are expected (together with BWDs) to populate the lower GW frequencies that will be probed by incoming space-based missions, such as LISA.
Keeping these astrophysical foregrounds under control is essential to our power of discovering a primordial GW background, which would enable us to test the physics describing primordial epochs which are complementary to what is testable through current probes
In the case of a causality-limited GW generation (such as a phase transition), this would open a window for precision cosmology through the accurate measurement of the low-frequency range of the spectrum \cite{Hook:2020phx}.
The knowledge of the astrophysical foregrounds is essential to this goal. In this paper, we consider them in detail in the frequency range of LISA, and we show that it is possible (with some assumptions that we detail in the following) to describe them with a linear model, where the unknown parameter is a prefactor multiplying a known spectral shape.
We then use the simple and numerically powerful formalism described in \cite{Poletti:2021ytu} to estimate the SNR by marginalising over the astrophysical foregrounds. 
We find that the sensitivity to a primordial signal that is not degenerate in spectrum with the astrophysical foregrounds remains promising.

There are three main assumptions that underlie this treatment for LISA, as illustrated in Sec.~\ref{sec:astro fg}.
Two of them concern the extrapolation of the BBH foreground as a $f^{2/3}$ power law from the $10-100\Hz$ range of LVK to the mHz range.
First, we need to confirm (as currently supported by e.g.~\cite{Bonetti:2020jku}) that the largest contribution to $\OGW^\bbh(f)$ at $f\sim\mHz-\Hz$ comes from the inspiral phase of lighter BHs merging in the LVK band, and not from heavier BHs (such as EMRIs) merging at lower frequencies. 
A second point of concern is the impact of eccentric BBHs on $\OGW^\bbh$, as $e$ is the only binary parameter that significantly modifies the amplitude $|h(f)|$ at $f\ll f_\textsc{isco}$.
Recent population synthesis studies \cite{Perigois:2020ymr, Sedda:2021vjh,Mapelli:2021gyv, Perigois:2021ovr} support the hypothesis that eccentric binaries do not alter $\OGW^\bbh\propto f^{2/3}$ above $\OGW\sim 10^{-13}-10^{-12}$. 
Future developments, regarding in particular the astrophysical modelling of dynamical channels for BBH formation, refinements on  BH and BBH population synthesis, and observational input from LVK, will reinforce these findings.
A third relevant astrophysical ingredient is the confusion foreground from BWDs which cannot be individually resolved. 
Recent studies for the galactic $\OGW^\bwd$, as \cite{Karnesis:2021tsh,Georgousi:2022uyt,Finch:2022prg}, confirm that this foreground will fall exponentially at $f>\mHz$ above the peak sensitivity of LISA, and future developments of population synthesis studies will refine our modelling of the residual foreground at lower frequencies.
Improvement in the estimation of the foreground from extragalactic BWDs is also required (see e.g.~\cite{Pan:2019uyn} for work in this direction), to assess its potential degeneracy with a weak primordial signal.

After having justified our modelling of the astrophysical uncertainties, we show how the sensitivity to cosmological parameters for precision cosmology is not dramatically impacted by the uncertainty on the foreground normalisation.
As visible in Figs.~\ref{fig:PT-w} and \ref{fig:PT-fFS}, the sensitivity to the equation-of-state parameter of the Universe $w$ or to the fraction of free-streaming particles $\fFS$ after a phase transition, worsen only by a factor of a few, and only where the signal is closer in shape and frequency to an astrophysical foreground.
This conclusion is guaranteed as long as the signal $\s$ is not degenerate in frequency space with an unknown astrophysical foreground $\a$, in the mathematical sense of $\s\cdot\a \ll \s\cdot\s$, and as long as the signal is the larger than the variance on the astrophysical normalisation, $\sa^2 \ll \s\cdot\s$ (see the definition of this scalar product as an integral over frequencies in Eq.~\eqref{eq:scalar product}, \eqref{eq:SNR astro}).

Our findings offer bright prospects for the detection of a primordial GW signal even in presence of astrophysical foregrounds, and at the same time highlight the importance of an accurate modelling of the astrophysical foregrounds from BBH and BWD mergers, since the knowledge of their spectral shape is essential to reduce their impact on the signal sensitivity.
A major reward of this programme is an unhindered potential to perform precision cosmology upon detection of a primordial GW signal.


\paragraph{Contributions.} 
D.R.\ lead the project, defined the analysis for all aspects related to the astrophysical and primordial signals, and wrote the manuscript. 
D.P.\ curated the foreground marginalization, the SNR methodology and their implementation, and contributed to the discussion of the results.

\acknowledgments
We thank Carole Périgois for useful discussions, and for providing us some numerical results from their paper \cite{Perigois:2021ovr}.
D.R.\ thanks 
Dawid Brzeminski,
Gabriele Franciolini, 
Anson Hook,
Junwu Huang,
Zhen Pan 
for useful discussions, and 
Valerio De Luca and Zhen Pan
for helpful comments on the draft.\\
D.R.\ is supported in part by NSF Grant PHY-2014215, DOE HEP QuantISED award \#100495, and the Gordon and Betty Moore Foundation Grant GBMF7946.
D.R.\ acknowledges hospitality from the Perimeter Institute for Theoretical Physics during the preparation of this paper.
Research at Perimeter Institute is supported in part by the Government of Canada through the Department of Innovation, Science and Economic Development Canada and by the Province of Ontario through the Ministry of Colleges and Universities.

 


\bibliographystyle{JHEP}
\bibliography{bib_SGWB-Discrimination}

\end{document}